\documentstyle[11pt,aaspp4]{article}
\def\lsim{\la}
\def\gsim{\ga}

\begin{document}

\title{\bf Radio Detection of Old GRB Remnants in the Local Universe}
\author{Eric Woods$^1$ and Abraham Loeb$^2$}
\medskip
\affil{Astronomy Department, Harvard University, 60 Garden St., Cambridge,
MA 02138} \altaffiltext{1}{email: ewoods@cfa.harvard.edu}
\altaffiltext{2}{email: aloeb@cfa.harvard.edu}

\begin{abstract}

We investigate the hydrodynamic evolution of Gamma--Ray Burst (GRB)
remnants at late times, $\gg 1~{\rm yr}$, by extending the standard
fireball model into the nonrelativistic Sedov--Taylor regime.  We calculate
the associated synchrotron luminosity as a function of remnant age. The
radio luminosity of old remnants depends strongly on the power--law index
$p\equiv d\log N/d \log\epsilon$ of the kinetic energy distribution of
shock-accelerated electrons.  In addition, the detection probability of
remnants in nearby galaxies depends on the GRB rate in the local universe.
We predict that if $p\la 2.6$ and the BATSE-calibrated GRB rate per
comoving volume is independent of redshift (despite the strong evolution in
the cosmic star formation rate) then the VLA can discover a $\sim 10^3$
years-old GRB remnant in the Virgo cluster. Nearby GRB remnants could be
searched for and resolved at radio frequencies, and then identified as
distinct from supernova remnants based on the detection of the extended
emission of high-ionization lines in follow--up optical observations.

\end{abstract}


\section{Introduction}

Since the detection of redshifted spectral features in gamma--ray burst
(GRB) afterglows, it has become evident that GRB events do indeed occur at
cosmological distances (Metzger et al. 1997; Djorgovski et al. 1998;
Kulkarni et al. 1998a; Bloom et al. 1998; Bloom et al. 1999).  The enormous
energy release $E\sim 10^{51}-10^{54}$ ergs implied by the observed
fluences and the cosmological distance scale could lead to a
relativistically--expanding fireball, which produces the prompt $(\lsim
100~{\rm sec})$ $\gamma$--ray emission due to collisions between its
internal shells (Paczy\'nski \& Xu 1994; Rees \& M\'esz\'aros 1994; Pilla
\& Loeb 1997; Kobayashi, Piran, \& Sari 1997). The fireball then enters the
afterglow phase at later times $(\gsim 1~{\rm day})$ when the expanding
wind decelerates due to its interaction with the surrounding medium
(e.g. M\'esz\'aros \& Rees 1997; Waxman 1997a, b).  During the early
afterglow phase, the blastwave is still ultra--relativistic, and is hence
well described by the self-similar Blandford--McKee solution (Blandford \&
McKee 1976). Within a month the expansion is no longer ultra--relativistic.
After about a year the expansion becomes nonrelativistic and the gas
dynamics is well approximated by the Sedov--Taylor self-similar solution
(Taylor 1950; Sedov 1959), similarly to a supernova remnant (SNR).  Given
the BATSE-calibrated rate of occurrence of GRBs in the universe (Wijers et
al. 1998) and the lifetime of their remnants, one infers that there should
be hydrodynamic fossils of GRB remnants in any spiral galaxy at any given
time (Loeb \& Perna 1998).  In fact, a subset of the so--called HI
supershells observed in the Milky Way and other nearby galaxies might be
old GRB remnants (Efremov et al. 1998; Loeb \& Perna 1998). Rhode et
al. (1999) have identified some HI holes in the nearby galaxy Holmberg II
that do not have optical counterparts as expected in alternative models
involving multiple SNe or starbursts with a normal stellar mass function.

There are two important physical differences which could in principle lead
to ways of observationally distinguishing GRB remnants from SNRs.  First,
the radiation energy emitted by an isotropic GRB explosion is estimated to
be up to 4 orders of magnitude higher than that released in a
supernova (Kumar 1999).  Hence if a nearby expanding remnant is discovered and the
explosion energy (as derived from the expansion speed and the Sedov-Taylor
solution) is too great to be attributable to a single supernova, then one
must conclude that we are seeing the remnant of either a GRB or multiple
supernovae.  But a GRB releases its total energy promptly, while the
explosion time of multiple supernovae in a star--forming region is not
expected to be synchronized to better than $\sim 10^6~{\rm years}$. Hence,
GRB remnants should be much more energetic than SN remnants at early times.
The existence of energetic X-ray remnants was recently inferred in deep
ROSAT X-ray images of M101 (Wang 1999); but future observations with the
XMM and Chandra X-ray satellites are necessary in order to identify the
nature of these sources conclusively through high-resolution imaging and
spectroscopy.  Second, GRB afterglows include a strong UV flash, which on
the timescale of a few hundreds of years creates an ionized bubble of
radius $\sim 100~n_1^{-1/3}~{\rm pc}$, where $n_1$ is the ambient density
in units of $1~{\rm cm}^{-3}$ (Perna \& Loeb 1998).  At ages $\la 10^4~{\rm
years}$, the non--relativistic shock wave acquires a much smaller radius
than the ionized bubble. This provides a potentially unique signature of
young GRB remnants; namely, a relatively compact expanding blastwave
embedded in a much larger ionized sphere.  The optical-UV recombination
lines from highly-ionized species in this region should provide a clear
discriminant for recognizing GRB remnants (Perna, Raymond \& Loeb
1999). The embedded blast wave emits both thermal and non-thermal
radiation; the latter being an extension of the afterglow due to
synchrotron emission by shock-accelerated electrons.

In this {\it Letter} we calculate the synchrotron radio emission from the
expanding GRB shock in old GRB remnants.  Our goal is to examine whether a
survey in the radio could efficiently identify old GRB remnants. Once a
candidate has been identified in the radio, follow-up optical observations
could search for the special recombination lines which are generic of GRB
remnants.  In \S 2 we review the blastwave hydrodynamics as extrapolated
into the Sedov--Taylor regime, and calculate the synchrotron flux as a
function of age, as well as the expected source counts in the Virgo
cluster.  Finally, we discuss our conclusions in \S 3.

\section{Hydrodynamics and Emission}

While the hydrodynamic evolution of GRB remnants through the early
afterglow phase is well-described by the ultra--relativistic
Blandford-McKee (1976) solution, at times much greater than a year we may
to a good approximation use the non-relativistic Sedov-Taylor solution.  It
has been shown (Huang, Dai, \& Lu 1998) that the two regimes may be
smoothly matched to one another.  We assume that the kinetic energy
remaining in the fireball is comparabale to the total energy $E_0$ which is
radiated away in $\gamma$-rays, and that the energy release is isotropic;
the effects of beaming in the initial energy release will be discussed in
\S 3.  The mean atomic weight of the surrounding (interstellar) medium is
taken to be $1.4m_p$, where $m_p$ is the proton mass.  The appropriate
scaling laws for the shock radius $r_s$ and velocity $v_s$ in the
non-relativistic regime are then
\begin{equation}
r_s = 1.17 \left(\frac{E_0 t^2}{1.4 m_p n}\right)^{0.2} = 1.56 \times
10^{18}~n_1^{-0.2}~E_{52}^{0.2}~t_{\rm yr}^{0.4}~{\rm cm},
\label{rsedov}
\end{equation}
\begin{equation}
v_s = 0.47 \left(\frac{E_0}{1.4 m_p n t^3}\right)^{0.2} = 1.99\times
10^{10}~n_1^{-0.2}~E_{52}^{0.2}~t_{\rm yr}^{-0.6}~{\rm cm~s^{-1}},
\label{vsedov}
\end{equation}
where $n=1~n_1~{\rm cm^{-3}}$ is the pre--shock particle number density,
$E_0=10^{52}~E_{52}~{\rm erg}$ is the total burst energy, and $t_{\rm yr}$
is the age of the remnant in years.

The state of the gas behind the shock is defined by the Rankine-Hugoniot
jump conditions.  For a strong shock, the post-shock (primed) particle
density, bulk velocity, and energy density are, respectively,
\begin{equation}
n^\prime = 4 n,
\label{njump}
\end{equation}
\begin{equation}
v^\prime = \frac{3}{4} v_s,
\label{vjump}
\end{equation}
\begin{equation}
u^\prime = \frac{9}{8} n m_p v_s^2,
\label{ujump}
\end{equation}
where all quantities are measured in the frame of the unshocked material.

As usual, we assume that the electrons are injected with a power--law
distribution of kinetic energies,
\begin{equation}
f(\epsilon) = (p-1)~\epsilon_m^{p-1}~\epsilon^{-p}, ~~~~~~~
\epsilon>\epsilon_m,
\label{powerlaw}
\end{equation}
where $f(\epsilon)d\epsilon$ is the fraction of shock-accelerated electrons
with kinetic energies in the range $(\epsilon,\epsilon+d\epsilon)$. We
assume a single power-law index at all energies, for simplicity.  The
spectral break that might exist at the transition to nonrelativistic
electron energies occurs well below the electron energy which is
responsible for the relevant radio emission at $\sim 1~{\rm GHz}$.

The post--shock magnetic and electron energy densities are assumed to be
given respectively as fixed fractions $\xi_b$ and $\xi_e$ of the total
post--shock energy density:
\begin{equation}
u^\prime_b = \frac{B'^2}{8\pi} = \xi_b u^\prime,
\end{equation}
\begin{equation}
u^\prime_e = n^\prime \epsilon_m~\frac{p-1}{p-2} = \xi_e u^\prime,
\end{equation}
where $m_e$ is the electron mass.
The magnetic field strength and minimum electron energy are thus given by
\begin{equation}
B^\prime = 0.14~\xi_b^{0.5}~n_1^{0.3}~E_{52}^{0.2}~ t_{\rm yr}^{-0.6}~{\rm
G},
\end{equation}
\begin{equation}
\epsilon_m = 117~\frac{p-2}{p-1}~\xi_e~n_1^{-0.4}~
E_{52}^{0.4}~t_{\rm yr}^{-1.2}~{\rm MeV}.
\end{equation}

Before calculating the flux, we need to determine whether the Sedov--Taylor
remnant is optically thin to synchrotron self--absorption.  The shocked
region has a thickness $\eta r_s$, with $\eta\sim 0.1$ from particle number
conservation.  We will adopt the value $\eta=1/15$, which is consistent
with the density profile of the Sedov (1959) solution, and provides a
number of shocked electrons which is a fair fraction of the total number
swept up by the blast wave.  The self--absorption optical depth at a photon
frequency $\nu$ is given by (cf. Rybicki \& Lightman 1979, p. 190)
\begin{equation}
\tau(\nu) = \frac{\sqrt{3}e^3nB_\perp \eta r_s}{2\pi m_e \epsilon_m \nu^2}~
(p-1)~\Gamma\left(\frac{p}{4}+\frac{11}{6}\right)
\Gamma\left(\frac{p}{4}+\frac{1}{6}\right)~
\left(\frac{\epsilon_m}{m_ec^2}\right)^p~ \left(\frac{3eB_\perp}{2\pi
m_ec\nu}\right)^{0.5p}.
\label{taunu}
\end{equation}
Here, $m_e$ is the electron mass, $\Gamma(x)$ is the gamma function, and
$B_\perp\approx B^\prime$ is the component of the magnetic field
perpendicular to the electron velocity. Substituting the numerical factors
into (\ref{taunu}) gives
\begin{equation}
\tau(\nu) = \frac{39.1\eta}{\nu_9^2}~
\left(\frac{61.6}{\nu_9}\right)^{0.5p}~\alpha(p)~
\xi_b^{0.5+0.25p}~\xi_e^{p-1}~n_1^{1.5-0.25p}~E_{52}^{0.5p}~ t_{\rm
yr}^{1-1.5p},
\end{equation}
where $\nu=10^9\nu_9~{\rm Hz}$, and
\begin{equation}
\alpha(p) \equiv (p-1)~\left(\frac{p-2}{p-1}\right)^{p-1}~
\Gamma\left(\frac{p}{4}+\frac{11}{6}\right)
\Gamma\left(\frac{p}{4}+\frac{1}{6}\right)
\end{equation}
is of order unity.  Thus, for typical power--law index values $2.1<p<3.2$
(Li \& Chevalier 1999), we find that $\tau(\nu) \ll 1$ for $\nu=1~{\rm
GHz}$ at an age of $\gg 1~{\rm yr}$.  

The flux from old remnants is therefore given by
\begin{equation}
F_\nu=\left(\frac{4\pi r_s^2 \cdot \eta r_s}{4\pi D^2}\right) \times P_\nu,
\end{equation}
where $D$ is the distance to the remnant, and $P_\nu$ is the volume
emissivity (in ${\rm erg~s^{-1}~cm^{-3}~Hz^{-1}}$), given by (Rybicki \&
Lightman 1979, p. 180)
\begin{equation}
P_\nu = \frac{4\sqrt{3}e^3nB_\perp}{m_ec^2}~\frac{p-1}{p+1}~
\Gamma\left(\frac{p}{4}+\frac{19}{12}\right)
\Gamma\left(\frac{p}{4}-\frac{1}{12}\right)~
\left(\frac{\epsilon_m}{m_ec^2}\right)^{p-1}~ \left(\frac{3eB_\perp}{2\pi
m_ec\nu}\right)^{0.5(p-1)}.
\label{pnu}
\end{equation}
Note that we are justified in using the synchrotron formula since the
electrons which emit at GHz frequencies are ultra--relativistic:
\begin{equation}
\frac{\epsilon}{m_e c^2} = \left(\frac{4\pi m_e c\nu}{3eB_\perp}\right)^{0.5}
 = 41.3~\nu_9^{0.5}~\xi_b^{-0.25}~n_1^{-0.15}~E_{52}^{-0.1}~t_{\rm yr}^{0.3}.
\label{syncheng}
\end{equation}
Coulomb collisions thermalize the electrons only at non-relativistic
energies, well below the energies of interest for this discussion.  Hence,
we are justified in using the power--law spectral shape in equation
(\ref{pnu}). The flux should eventually show a cooling break at high
frequencies, due to the fact that electrons with more than a threshold
energy $\epsilon_c$ will radiate away their energy faster than the
dynamical time.  The synchrotron cooling time is given by $t_c = 6\pi(m_e
c^2)^2/\sigma_T c\epsilon B_\perp^2$, where $\sigma_T$ is the Thomson cross
section.  The cooling frequency is thus given by
\begin{equation}
\nu_c = \left(\frac{3eB_\perp}{4\pi m_e c}\right)~
\left(\frac{6\pi m_e c}{\sigma_T B_\perp^2 t}\right)^2~
= 9.4 \times 10^{11}~\xi_b^{-1.5}~n_1^{-0.9}~E_{52}^{-0.6}~t_{\rm yr}^{-0.2}~
{\rm Hz}.
\label{cooling}
\end{equation}
For remnant ages $\la 10^7~{\rm yr}$, the cooling cutoff occurs well above
1 GHz.  Thus, the emission at GHz frequencies should fall on the simple
$P_\nu\propto\nu^{(1-p)/2}$ part of the spectrum.

We may now calculate the numerical value of the expected flux 
\begin{equation}
F_\nu = \frac{5.2\times 10^6\eta}{D_{\rm Mpc}^2}~
\left(\frac{61.6}{\nu_9}\right)^{0.5(p-1)}~\beta(p)~
\xi_b^{0.25(1+p)}~\xi_e^{p-1}~n_1^{0.95-0.25p}~ E_{52}^{0.3+0.5p}~t_{\rm
yr}^{2.1-1.5p}~{\rm Jy},
\label{fnu}
\end{equation}
where $D_{\rm Mpc}$ is the distance to the source in Mpc, and
\begin{equation}
\beta(p) \equiv \left(\frac{p-1}{p+1}\right)~
\left(\frac{p-2}{p-1}\right)^{p-1}~
\Gamma\left(\frac{p}{4}+\frac{19}{12}\right)
\Gamma\left(\frac{p}{4}-\frac{1}{12}\right)
\end{equation}
is of order unity.

In figure 1 we plot the flux at 1.6 GHz from equation (\ref{fnu}) as a
function of remnant age $t_{\rm yr}$ for $p=2.1$ ($F_\nu\propto t^{-1.05}$)
and $p=3.2$ ($F_\nu\propto t^{-2.7}$), the two values which bracket the
range of power--law indices seen in radio supernovae (Li \& Chevalier
1999).  We have assumed that the sources reside in the Virgo cluster, at a
distance of $D_{\rm Mpc}=16$. We also assumed sub-equipartition energy
density of the magnetic fields and nonthermal electrons in the post-shock
gas with $\xi_b=0.1$, $\xi_e=0.1$; a typical interstellar medium density,
$n_1=1$; and $\eta=1/15$.  In panel (a), we take $E_{52}=0.7$, derived from
the BATSE catalog for a source population with a redshift--independent
burst rate per comoving volume, while in panel (b) we use $E_{52}=14$,
which applies for population which traces the star formation rate in
galaxies.  For comparison, the horizontal line shows the $5\sigma$
sensitivity threshold, $F_{\rm VLA}=70 \mu{\rm Jy}$, for the Very Large
Array (VLA) in the 1.6 GHz band, using a one--hour integration in the VLA's
most extended configuration (L. Greenhill 1999, private communication).
Clearly the maximum age for a detectable remnant, and hence the number of
detectable remnants, depends strongly on $p$.  The vertical line, at
$t=600$ yr and $10^5$ yr in panels (a) and (b) respectively, indicates the
age at which we expect to detect at least one remnant in the Virgo cluster.
The intermediate plots of $F_\nu(t)$ in panels (a) and (b) are for $p=2.6$
and 2.3, the respective maximum electron slopes for which a 600 or
$10^5$--year--old Virgo remnant would be detectable by the VLA.

To calculate the expected source counts, we note that the local GRB rate
per $L_\star$ (where $L_\star$ is the characteristic stellar luminosity per
galaxy in the local universe), assuming no beaming, is estimated to be
$\Gamma\approx 2.5\times 10^{-8}~L_\star^{-1}~{\rm yr^{-1}}$ (Wijers et al
1998).  This value is derived assuming that the GRB rate traces the star
formation history of galaxies; for a non--evolving burst rate, the inferred
value is 150 times higher.  The number of remnants per $L_\star$ younger
than $t_{\rm yr}$ [and hence brighter than $F_\nu(t_{\rm yr})$] at a given
time is then $\Gamma t_{\rm yr}$.  The best place to search for {\it
optical} emission from old GRB remnants is in the Virgo cluster (Perna,
Raymond, \& Loeb 1999).  There are $\sim 2500$ galaxies brighter than
$B=19$ in this cluster; at a distance of 16 Mpc, this limit corresponds to
an absolute magnitude magnitude $M_B=-12$.  For typical Schechter (1976)
function parameters, $\alpha=-1$ and $M_\star=-19$ (Loveday et al. 1992;
Marzke et al. 1994), this yields a total luminosity for the Virgo cluster
of $L_{\rm Vir}=430 L_\star$.  Thus, for a non-evolving GRB population, we
need to look for remnants as old as $t=(\Gamma\cdot L_{\rm Vir})^{-1}=600$
yr to be reasonably confident of seeing at least one remnant in the Virgo
cluster.  For a GRB population which traces the cosmic star formation
history, we need $t=10^5$ yr.  As illustrated in Figure 1, for the burst
parameters we have assumed, remnants this old could be detected by the VLA
only for an electron power--law index $p<2.6$ in the non--evolving case and
$p<2.3$ in the evolving case.

For deeper volume--limited searches, e.g. of volumes probed by the Sloan
Digital Sky Survey (SDSS; Gunn \& Weinberg 1995), the increase in the
number of galaxies surveyed $N\propto D^3$ dominates over the decrease in
detectable ages $t\propto D^{2/(2.1-1.5p)}$, obtained by solving equation
(\ref{fnu}) for $t_{\rm yr}$.  Thus, although at a distance of 250 Mpc we
need to look for remnants which are younger than $10^2$ years, the galaxy
luminosity within this volume is $\sim 10^5 L_\star$, leading to the
prediction of at least one detectable young remnant for $p=3.2$ and many
more for lower $p$ values.

\section{Discussion and Conclusions}

We have calculated the expected synchrotron flux from $\la 10^5$--year--old
GRB remnants in the Virgo cluster.  Although the most revealing signature
of a GRB remnant may be the presence of optical-UV recombination lines from
high-ionization states of metals (Perna, Raymond, \& Loeb 1999), it might
be easier to search for nearby GRB remnants in the radio due to the lower
background noise (from the sky plus the host galaxy) at radio frequencies.
For a non--evolving GRB population, we find that if the spectral index of
the shock-accelerated electrons $p<2.6$, then one could find at least one
$\sim 600$--year--old remnant in the Virgo cluster at the VLA detection
threshold. Such a remnant should be characterized by a synchrotron spectrum
$F_\nu \propto \nu^{-0.8}$ or flatter.
\footnote{Although it appears that the energy flux $\nu F_\nu \propto
\nu^{0.2}$ diverges at high frequencies, recall (cf. Eq.
\ref{cooling}) that there is a break in the spectrum at the cooling
frequency $\nu_c$, above which, $F_\nu \propto \nu^{-0.3}$.}

Perhaps the most poorly constrained parameter of the GRB sources is the
beaming fraction $f_b$, the fraction of $4\pi$ steradians into which the
initial $\gamma$--ray emission is emitted; the actual event rate may then
be enhanced to $f_b^{-1}\Gamma$. However, the total energy released in an
event scales as $f_b E_0$, and so the synchrotron flux from the remnant is
proportional to $f_b^{0.3+0.5p}$.  The decrease in the required energy due
to beaming may be counteracted by the fact that the efficiency for
producing $\gamma$--rays in the initial event is very low (Kumar
1999).  We note that even if the initial (impulsive) energy release is
beamed, the deposited energy, $f_b E_0$, will be isotropized at the onset
of the non-relativistic expansion phase. In addition, for a uniform
distribution of sources in Eucledean space, the number of detectable
remnants younger than a given age should decline with decreasing $f_b$ as
$\propto f_b^{-1} (f_b^{0.3+0.5p})^{3/2}= f_b^{0.75p-0.65}$.


Due to the as--yet uncertain nature of the progenitors, it is not clear
whether we should expect the GRB rate to directly trace the star formation
history.  Hence we considered both the SFR--tracing and non--evolving cases
in Figure 1.  We summarize these results in Figure 2, which shows the upper
bound on $p$ as a function of explosion energy for the two cases.  The
factor of 150 enhancement in the rate for a non--evolving population allows
for a greater probability of seeing younger bursts with steeper spectral
slopes.  It is important to note that if a young $(\lsim 10^3~{\rm yr})$
remnant is detected in the Virgo cluster, our results will strongly suggest
that the local GRB rate is consistent with the non--evolving scenario, but
not with the star--formation--tracing scenario.  This is the only immediate
way to constrain the local burst rate (short of monitoring the $\sim 10^6$
SDSS galaxies over one year and searching for a GRB explosion).  The young
radio remnants detected in this case should be well--embedded in the bubble
that was ionized by the initial UV flash, and should be detectable in
optical recombination lines.  If GRBs follow the global star formation
history, however, we only expect to see a $\sim 10^5$--year--old remnant in
Virgo, which is large and not well-embedded, and will also not be easily
detectable at optical wavelengths (Perna, Raymond, \& Loeb 1999).

Is it justified to assume that the only significant flux at 1.6 GHz comes
from the freshly shocked electrons within a distance $\eta r_s$ behind the
shock? Might the electrons which were shocked at $t\sim 1~{\rm yr}$
contribute a substantial amount of flux when the remnant is $10^4$ years
old?  The Sedov-Taylor similarity solution implies that the volume occupied
by the material behind the shock front is increasing with time as $r_s^3$.
Hence, the relativistic energy densities of the electrons and magnetic
fields scale adiabatically as $u_b'\propto u_e' \propto r_s^{-4}$.  The
flux at a fixed frequency $\nu$ due to the old electrons is then
$F_\nu^{\rm old}\propto t^{-0.8p_{\rm old}}$, where $p_{\rm old}$ is the
power--law index measured for afterglows at $\sim 1~{\rm yr}$; typically
$p_{\rm old}\approx 3$.  Using this value, a comparison with equation
(\ref{fnu}) yields a ratio $F_\nu^{\rm old}/F_\nu^{\rm new} = t^{1.5p_{\rm
new}-4.5}$, where $F_\nu^{\rm new}$ is the flux from newly--shocked
electrons.  Thus, if $p_{\rm new}>3$, the flux from the old electrons will
dominate at times $t\gg 1~{\rm yr}$. For values in the range observed in
radio SNRs (Li \& Chevalier 1999), we are justified in neglecting the
contribution from the old electrons.

Our extrapolation to the non-relativistic regime may be tested by
continuing to monitor the event GRB 970508, which is at $z=0.835$ and had a
1.4 GHz flux of $249\pm 60~\mu{\rm Jy}$ at an age of 354 days (Frail et
al. 1999, in preparation).

Our discussion assumed that GRB sources release their energy impulsively in
the form of the ultra-relativistic wind that produces the early
$\gamma$-ray and afterglow emission.  It is possible that more energy is
released in the form of non-relativistic ejecta that catches-up with the
decelerating shock and re-energizes it at late times. In this case, if
$f_b=1$ then the total hydrodynamic energy (and the associated synchrotron
flux) may be much larger than we estimated based on the GRB energetics
alone. However, the situation might be different if $f_b\ll 1$.  Recently,
there have been several claims for a potential detection of supernova
emission in the light curves of rapidly declining afterglows (Kulkarni et
al. 1998b; Bloom et al. 1999; Reichart 1999). We emphasize that any
association between supernova and GRB events could be explored more
directly in the local Universe by examining the ionization and hydrodynamic
structure of SN--like remnants in the interstellar medium of nearby
galaxies.  For example, one could search for extended ionization cones in
young supernova remnants, as expected if the intense UV emission from the
associated GRB afterglows is collimated.  Complementary information about
the shock structure and temperature can be obtained by observations in the
radio band (probing the synchrotron emission) or the X-ray regime (probing
thermal emission from the hot post-shock gas).  In particular, follow-up
observations of the energetic X-ray remnants discovered by Wang (1999) in
M101 or the optically-faint HI holes discovered by Rhode et al. (1999) in
Holmberg II, would be revealing as to the nature of these peculiar objects.

The expanding spherical shock front will typically acquire a diameter of
$\sim 1^{\prime\prime}$ after $10^4$ years at the distance of the Virgo
cluster.  With VLBI, one can achieve up to sub milli-arcsecond resolution
and hence easily resolve the radio--emitting shock. Since the shocked gas
occupies a thin shell, the source should be strongly limb-brightened and
possibly highly polarized at the limb (Medvedev \& Loeb 1999).  In the
simplest case of an isotropic point explosion, the radio--emitting region
will appear embedded inside a much larger region which was ionized by the
prompt UV emission from the GRB and which emits optical-UV recombination
lines (Perna, Raymond, \& Loeb 1999).  This distinct structure of a shock
embedded in an HII region with high ionization states of heavy elements, is
unique to GRB remnants since only the optically-thin wind of a relativistic
GRB fireball can give rise to the intense hard radiation which produces a
highly-ionized bubble out to large distances $\sim 100~{\rm pc}$, in front
of the shock.  This is to be contrasted with ordinary supernovae, in which
the UV emission from the optically-thick envelope is suppressed above the
thermal cutoff.

\acknowledgments

We thank Lincoln Greenhill for useful discussions. 
This work was supported in part by NASA grants NAG 5-7039 and NAG 5-7768.

\vfill\eject
\begin{figure}
\plotone{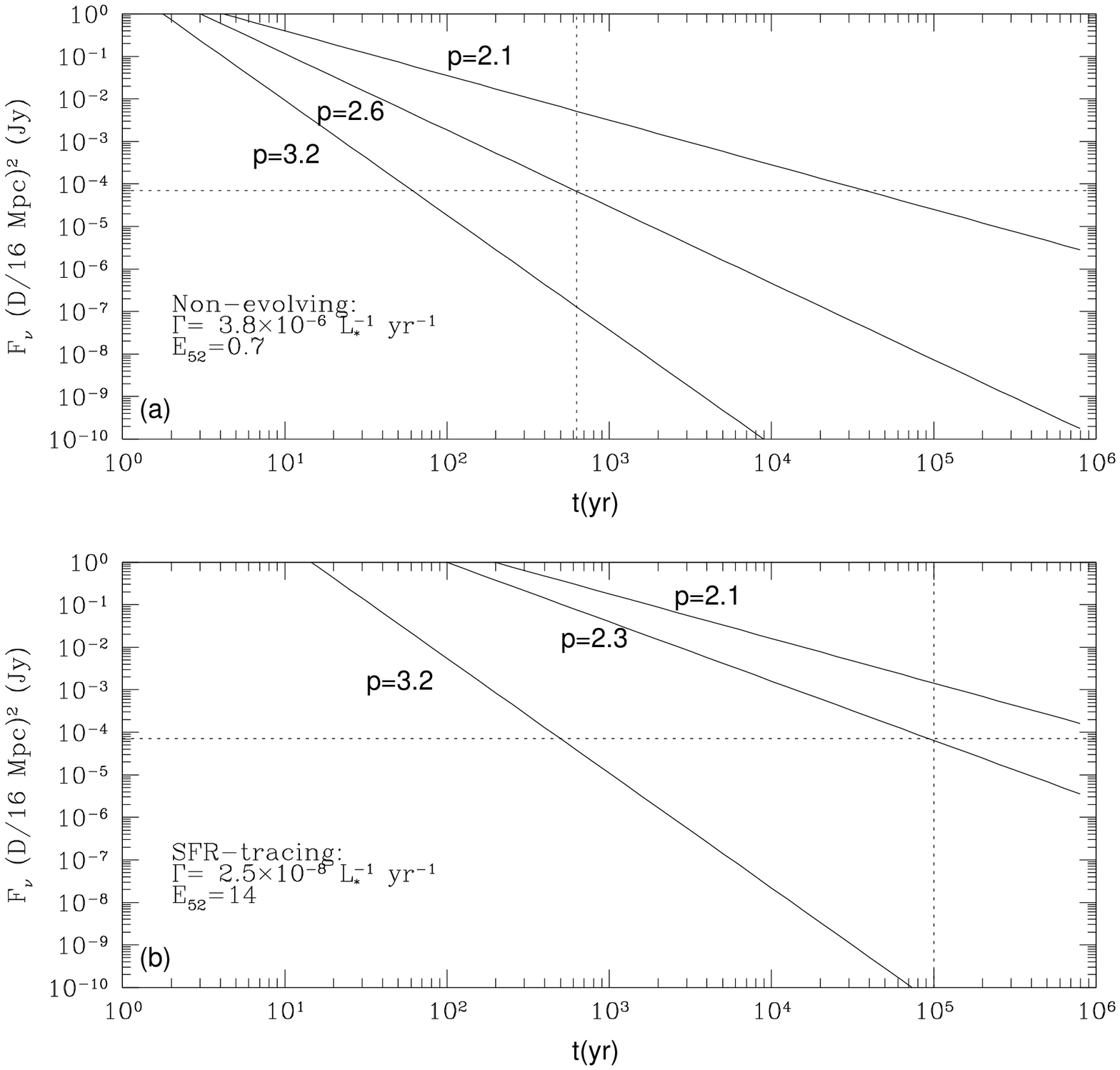}
\caption{Observed remnant 1.6~GHz flux as a function of age (solid lines)
at the Virgo cluster distance $D=16$ Mpc, for (a) a non--evolving GRB
population and (b) a star--formation--tracing GRB population.  We show
three different power--law indices; the upper and lower values correspond
to the range of observed slopes in radio SNRs, while the middle value is
such that one remnant should be visible in the Virgo cluster.  Shown for
comparison is the VLA $5\sigma$ sensitivity level, $F_{\rm VLA}=70~\mu{\rm
Jy}$ (horizontal dotted line), and the corresponding age limit [$t=600$ yr
in panel (a) and $t=10^5$ yr in panel (b)], such that there should be at
least one younger remnant in Virgo.  We assume $\xi_b=\xi_e=0.1$, and
$\eta=1/15$.}
\end{figure}

\vfill\eject
\begin{figure}
\plotone{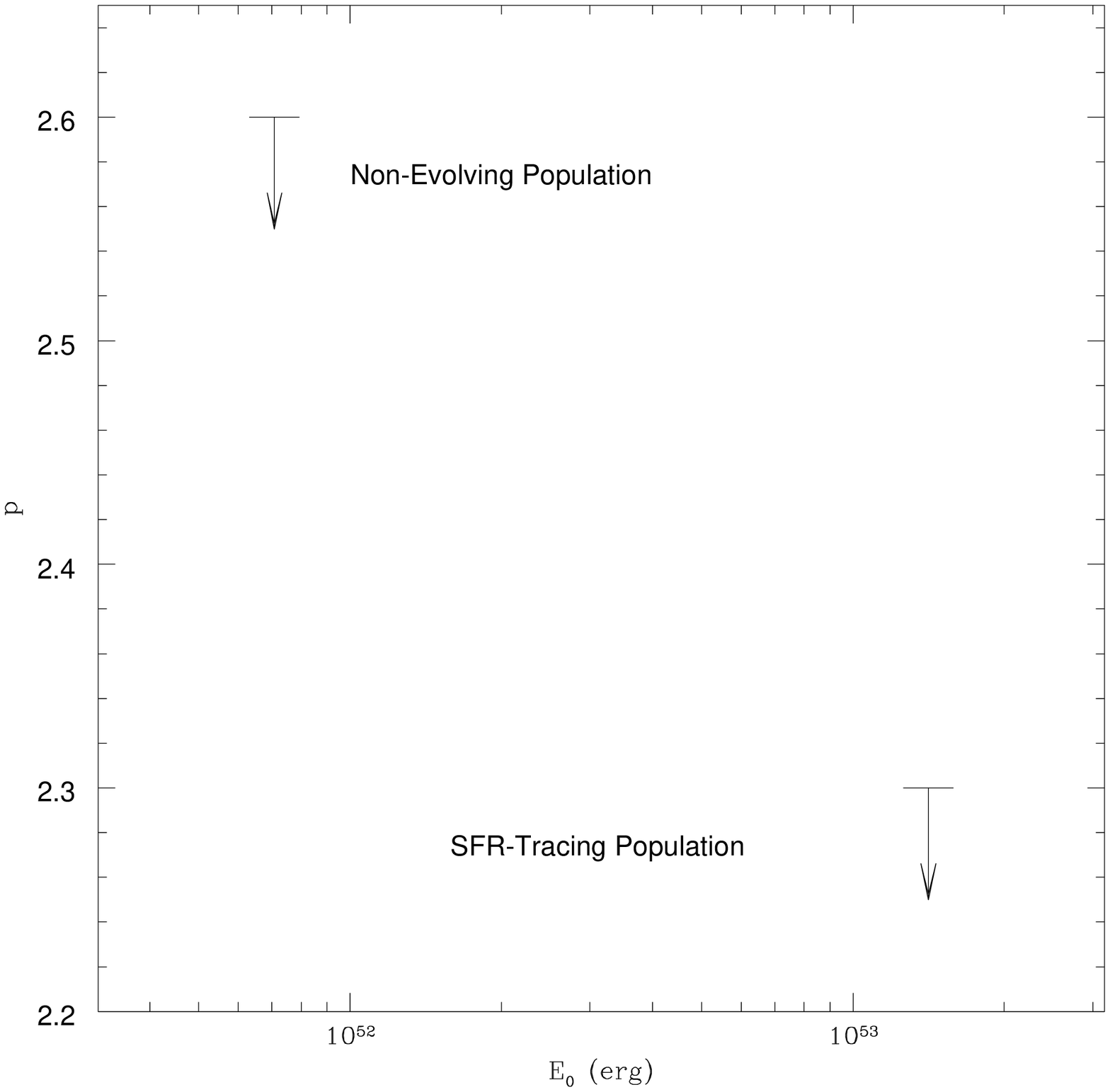}
\caption{Maximum values of the electron power--law slope $p$, such that one
remnant will be detectable in the Virgo cluster (assuming $f_b=1$).  We
show best--fit results for a non--evolving source population (upper left)
and a population which traces the star formation history of galaxies (lower
right).  The widths of the horizontal line segments reflect the $1\sigma$
uncertainty in the burst energy, from Wijers et al. (1998).}
\end{figure}


\begin{references}

\reference{}
Blandford, R. D., \& McKee, C. F. 1976, Phys. Fluids, 19, 1130

\reference{}
Bloom, J. S., Fenimore, E. E., \& in't Zand, J. 1996, in Proc. of
the Huntsville Symposium on GRBs (New York: AIP), 321

\reference{}
Bloom, J. S., et al. 1999, ApJ, 518, L1

\reference{}
Bloom, J. S., et al. 1999, Nature, in press, astro-ph/9905301

\reference{}
Djorgovksi, S. G., et al. 1998, ApJ, 508, L17

\reference{} 
Efremov, Y. N., Elmegreen, B. G., Hodge, P. W. 1998, ApJ, 501, L163

\reference{}
Gunn, J. E., \& Weinberg, D. H. 1995, in Wide Field Spectroscopy
and the Distant Universe, eds. S. Maddox \& A. Arag\'on--Salamanca
(Singapore: World Scientific), p. 3

\reference{}
Huang, Y. F., Dai, Z. G., \& Lu, T. 1998, A\&A, 336, L69

\reference{}
Kobayashi, S., Piran, T., \& Sari, R. 1997, ApJ, 490, 92

\reference{}
Kulkarni, S. R., et al. 1998a, Nature, 393, 35

\reference{}
Kulkarni, S. R., et al. 1998b, Nature, 395, 663

\reference{}
Kumar, P. 1999, ApJL, in press, astro-ph/9907096

\reference{}
Li, Z., \& Chevalier, R. A. 1999, ApJ, in press, astro-ph/9903483

\reference{}
Loeb, A., \& Perna, R. 1998, ApJ, 503, L35

\reference{}
Loveday, J., Peterson, B. A., Efstathiou, G., \& Maddox, S. J.
1992, ApJ, 390, 338

\reference{}
Mahadevan, R., Narayan, R., \& Yi, I. 1996, ApJ, 465, 327

\reference{}
Marzke, R. O., Geller, M. J., Huchra, J. P., \& Corwin, H. G.
1994, AJ, 108, 437

\reference{}
Medvedev, M. V., \& Loeb, A. 1999, ApJ, in press, astro-ph/9904363

\reference{}
M\'esz\'aros, P., \& Rees, M. J. 1997, ApJ, 476, 232

\reference{}
Metzger, M. R., et al. 1997, Nature, 387, 879

\reference{}
Ostriker, J. P, \& McKee, C. F. 1988, Rev. Mod. Phys., 60, 1

\reference{}
Paczy\'nski, B., \& Xu, G. 1994, ApJ, 427, 708

\reference{}
Perna, R., \& Loeb, A. 1998, ApJ, 501, 467

\reference{}
Perna, R., Raymond, J., \& Loeb, A. 1999, ApJ, submitted,
astro-ph/9904181

\reference{} Pilla, R. P., \& Shaham, J. 1997, ApJ, 486, 903

\reference{} Pilla, R. P., \& Loeb, A. 1998, ApJ, 494, L167

\reference{}
Rees, M. J., \& M\'esz\'aros, P. 1994, ApJ, 430, L93

\reference{}
Reichart, D. E. 1999, ApJL, in press, astro-ph/9906079

\reference{} Rhode, K. L., Salzer, J. J., Westfahl, D. J., \& Radic,
L. 1999, AJ, in press, astro-ph/9904065

\reference{}
Rybicki, G. B., \& Lightman, A. P. 1979, Radiative Processes in
Astrophysics (New York: Wiley--Interscience)

\reference{}
Schechter, P. 1976, ApJ, 203, 297

\reference{}
Sedov, L. I. 1959, Similarity and Dimensional Methods in Mechanics
(New York: Academic)

\reference{}
Taylor, G. I. 1950, Proc. Roy. Soc. London A, 201, 159

\reference{}
Wang, D. 1999, ApJL, in press, astro-ph/9903246

\reference{}
Waxman, E. 1997a, ApJ, 485, L5

\reference{}
Waxman, E. 1997b, ApJ, 489, L33

\reference{}
Wijers, R. A. M. J., Bloom, J. S., Bagla, J. S., \& Natarajan, P.
1998, MNRAS, 294, 13

\end{references}
\end{document}